\documentclass{Interspeech2024}

\interspeechcameraready

\title{SaSLaW: Dialogue Speech Corpus with Audio-visual Egocentric Information Toward Environment-adaptive Dialogue Speech Synthesis}

\name[affiliation={1}]{Osamu}{Take}
\name[affiliation={1,2}]{Shinnosuke}{Takamichi}
\name[affiliation={1}]{Kentaro}{Seki}
\name[affiliation={3}]{Yoshiaki}{Bando}
\name[affiliation={1}]{Hiroshi}{Saruwatari}

\address{
  $^1$The University of Tokyo, Japan,
  $^2$Keio University, Japan \\
  $^3$National Institute of Advanced Industrial Science and Technology (AIST), Japan}
\email{flymoons0325@g.ecc.u-tokyo.ac.jp}


\begin{document}

\maketitle

\begin{abstract}
    This paper presents \textit{SaSLaW}, a spontaneous dialogue speech corpus containing synchronous recordings of what speakers speak, listen to, and watch. Humans consider the diverse environmental factors and then control the features of their utterances in face-to-face voice communications. Spoken dialogue systems capable of this adaptation to these \textit{audio environments} enable natural and seamless communications. SaSLaW was developed to model human-speech adjustment for audio environments via first-person audio-visual perceptions in spontaneous dialogues. We propose the construction methodology of SaSLaW and display the analysis result of the corpus. We additionally conducted an experiment to develop text-to-speech models using SaSLaW and evaluate their performance of adaptations to audio environments. The results indicate that models incorporating hearing-audio data output more plausible speech tailored to diverse audio environments than the vanilla text-to-speech model.
\end{abstract}

\vspace{-1mm}
\section{Introduction} \vspace{-1mm}
Text-to-speech (TTS) is an important technology for spoken dialogue systems (SDSs) such as conversational robots~\cite{kawahara2019iwsds}. 
These systems are often implemented in conversational scenarios within real-world environments. 
Human-to-human voice communication in real environments often involves natural and intelligible speech tailored to surrounding factors such as background noise and their physical proximity. We call these environmental factors collectively as the \textit{audio environment}.

\begin{figure}[t]
  \centering
  \includegraphics[width=\linewidth]{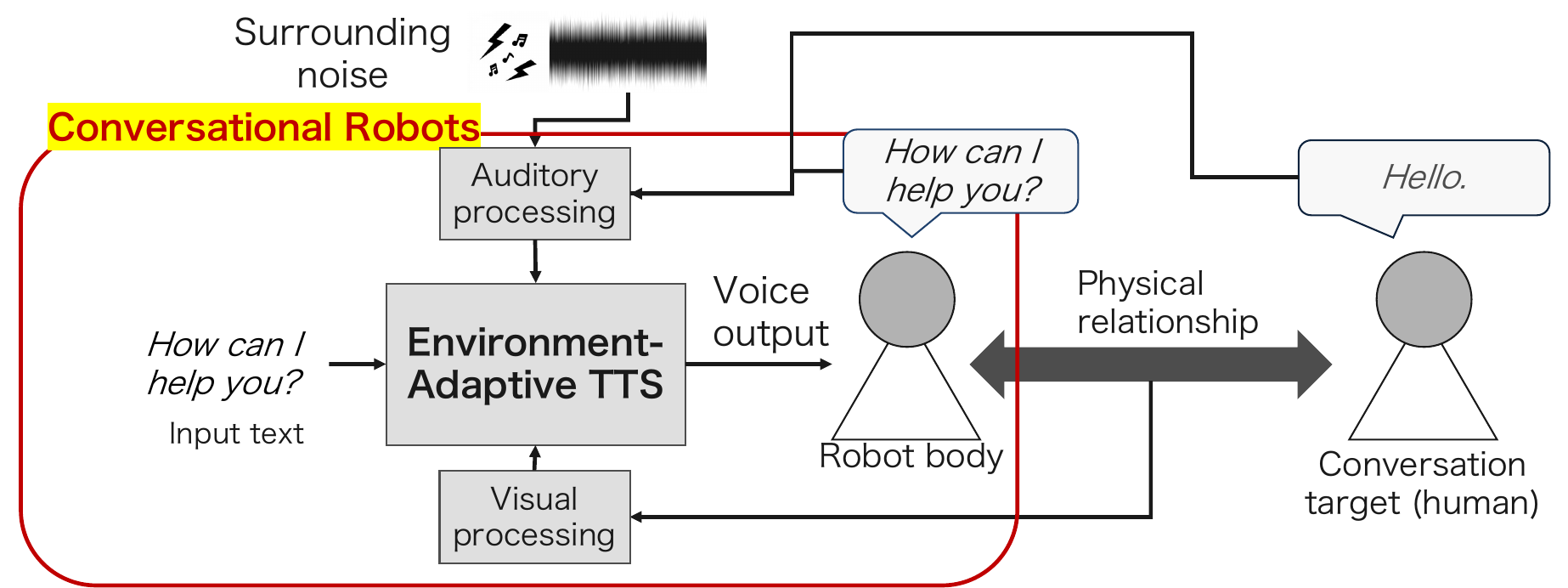}
  \vspace{-4mm}
  \caption{Dialogue agent that generates speech using environment-adaptive TTS (EA-TTS). EA-TTS changes the speech style affected by interlocutor or environmental noise, just like us humans. This paper proposes a methodology of corpus construction to realize this and build open-source corpus.}
  \vspace{-3mm}
  \label{fig:robot-with-eatts}
\end{figure}

The adaptation to audio environments by humans is based on the auditory and visual information they perceive~\cite{cooke2014csl, Hazan2011AcousticphoneticCO}, which can be explained within the framework of the speech chain~\cite{denes1993speech}.
Reports have also indicated that different audio environments necessitate natural speech variations of conversational robots for humans~\cite{tuttosi2023iros}. Therefore, TTS incorporating audio environment inputs with the framework of speech chain is necessary for SDSs to achieve natural and seamless speech communication in dialogues. We refer to this dialogue TTS as environment-adaptive TTS (EA-TTS). Figure~\ref{fig:robot-with-eatts} illustrates the application of EA-TTS.

Deep neural networks~(DNNs) and prevailing large-scale corpora~\cite{neuraltts} enable TTS models to generate natural speech comparable to humans for read speech in quiet environments.
However, an EA-TTS model cannot be constructed only with the speaker's clean speech recorded in quiet backgrounds. EA-TTS should require first-person recordings of what humans speak, hear, and see during dialogues in various audio environments. Corpora for such EA-TTS and their construction methods are yet to be established despite the prevalence of TTS corpora.

We present a methodology to construct a spontaneous dialogue speech corpus containing synchronous recordings of egocentric audio-visual perceptions. Following this methodology, a novel speech corpus called \textit{SaSLaW}\footnote{``\textbf{S}o, what \textbf{a}re you \textbf{S}peaking, \textbf{L}istening, \textbf{a}nd \textbf{W}atching?''} is constructed and published\footnote{\url{https://github.com/sarulab-speech/SaSLaW}}. We also construct EA-TTS models based on DNNs using SaSLaW and conduct a comparative evaluation. 

\vspace{-1mm}
\section{Related Work} \vspace{-1mm}

\subsection{Emulating Adaptations to Audio Environments} \vspace{-1mm}
\textbf{Noise environment.} The Lombard effect~\cite{lombard} describes the involuntary voice raising of humans in noisy environments.
Several methods have been proposed for mimicking the Lombard effect, including using signal-processing-inspired manipulations~\cite{zorila12interspeech, Thuan2020ieice} and neural TTS~\cite{bollepalli19_interspeech, novitasari2022taslp, raitio22_interspeech}.
However, previous research~\cite{raitio22_interspeech} confirmed the degradation in the naturalness of manipulated speech by signal processing.
These methods are also limited to the study of read speech and do not model the Lombard effect in spontaneous dialogue~\cite{folk2011interspeech}.

\textbf{Toward listeners.} Prosody of human speech is also affected by the physical proximity between talkers~\cite{pelegrin2011jasa} and interlocutors' speech~\cite{nishimura09jpsj}, a phenomenon known as \textit{entrainment}~\cite{levitan-etal-2012-acoustic}. 
Several methods construct neural TTS models with architectures for adapting to interlocutors' speech~\cite{Li2023acmicm} and faces~\cite{zhou2023icassp}. However, none of these methods address both spontaneous dialogue scenes and the use of egocentric perceptions.

\vspace{-1mm}
\subsection{Corpora for Environment-adaptive TTS} \vspace{-1mm}
Several corpora~\cite{Cooke013interspeech, alghamdi2018jasa} have been constructed for TTS with modeling adaptations to noisy surroundings.
These corpora primarily focus on modeling read-style Lombard speech. The construction of TTS corpora modeling adaptation to environmental noises in spontaneous conversations remains unexplored.

There are TTS corpora for modeling dialogue scenes used with DNN-based TTS models~\cite{guo2021slt, koiso-etal-2022-design}. The CEJC corpus~\cite{koiso-etal-2022-design} contains spontaneous conversations in real environments, while capturing environmental noise and visual footage from a third-person perspective.
The EasyCom~\cite{donley2021easycom} dataset contains participants' speech and egocentric videos during conversations in noisy environments but lacks the variety of audio environments. EgoCom~\cite{northcutt2023tpami} contains firsthand audio-visual experiences but lacks speech recordings with minimal external noise as it was primarily designed for recognition and understanding tasks.
Table~\ref{tab:corpus-comparison} compares these corpora to our SaSLaW.
\begin{table}[tb] 
    \caption{Corpora comparison. ``Spon.'' and ``perf.'' are spontaneous and performative styles, respectively. ``fp'' and ``tp'' are first- and third-persons views, respectively. ``IR'' indicates impulse responses between talkers, and  ${}^\dag$ means near-real noise.} 
    \label{tab:corpus-comparison}
    \vspace{-3mm}
    \footnotesize
    \hbox to\hsize{\hfil
    \begin{tabular}{lcccccc}\hline\hline
        corpus                                      & style & noise         & hear  & see   & speak & IR\\\hline
        \multicolumn{3}{l}{\textbf{TTS corpus}} \\
        \textbf{SaSLaW(ours)}                       & spon. & real${}^\dag$ & fp    & fp    & \checkmark    & \checkmark \\
        Hurricane~\cite{Cooke013interspeech}        & read  & real${}^\dag$ & -     & -     & \checkmark    & -\\
        CEJC~\cite{koiso-etal-2022-design}          & spon. & real          & tp    & tp    & \checkmark    & -\\
        Guo et al.~\cite{guo2021slt}                 & perf. & -             & -     & -     & \checkmark    & -\\\hline
        \multicolumn{3}{l}{\textbf{Datasets not focusing on TTS}} \\
        EgoCom~\cite{northcutt2023tpami}            & spon. & real          & fp    & fp    & -             & -\\
        EasyCom~\cite{donley2021easycom}            & spon. & real          & tp    & fp    & \checkmark    & - \\
        Hurricane 2.0~\cite{rennies20interspeech}   & read  & real${}^\dag$ & tp    & -     & \checkmark    & \checkmark\\\hline
    \end{tabular}\hfil}
\end{table}


\vspace{-1mm}
\section{Corpus-construction Methodology} \vspace{-1mm}
\label{sec:corpus-construction}
We describe the construction methodology of the spontaneous dialogue corpus SaSLaW, which includes first-person-perspective multi-modal information.

\vspace{-1mm}
\subsection{Overview of SaSLaW} 
\begin{figure}[t]
  \centering
  \includegraphics[width=\linewidth]{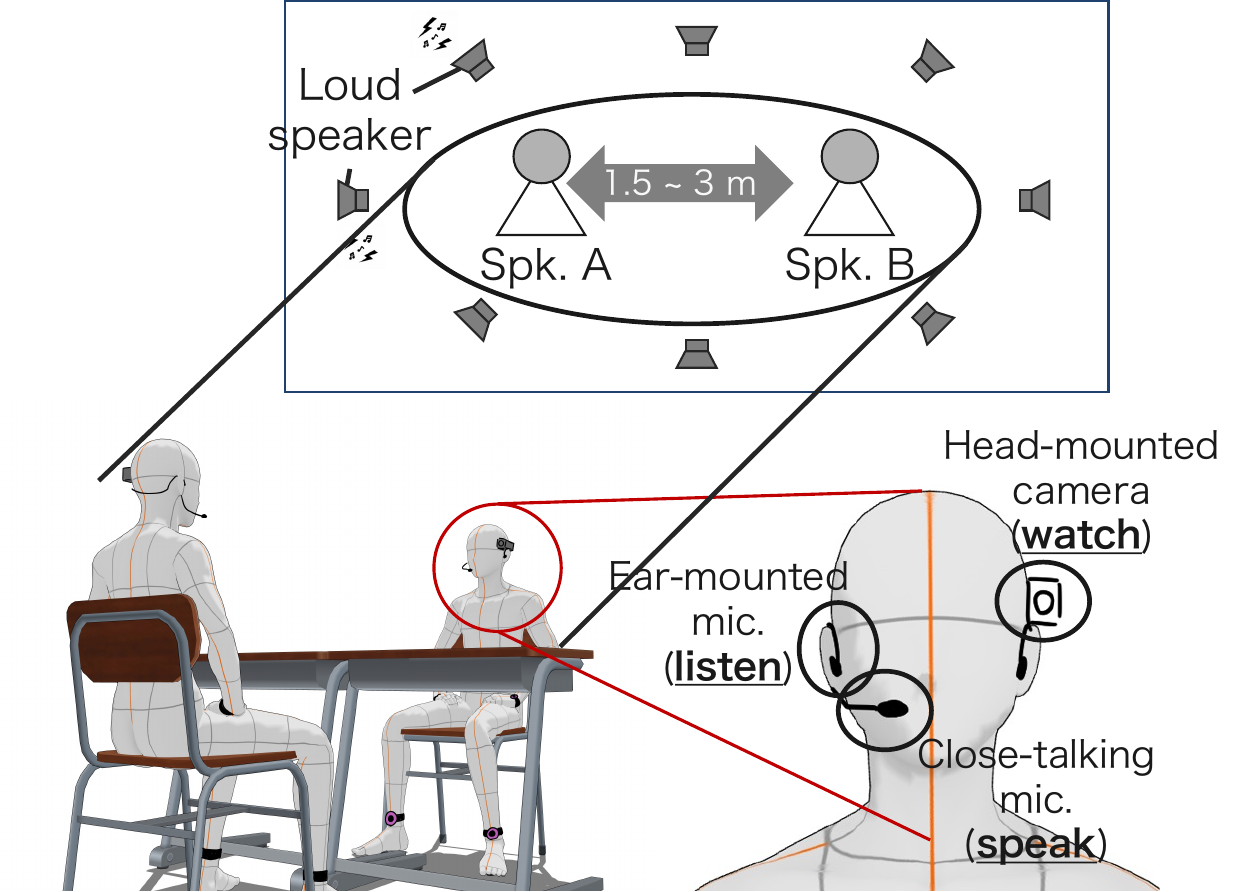}
  \caption{Recording configuration during two-person conversation. Top illustrates the configuration of the noisy environment, and bottom illustrates participants' equipment.}
  \label{fig:saslaw-record-info}
\end{figure}

\vspace{-1mm}
SaSLaW captures adaptation to audio environments (noise, interlocutor) in spontaneous human speech communication. Achieving his involves recording scenes in which two participants engage in spontaneous dialogue while facing each other in a simulated noisy environment, mimicking real-world conditions.
Figure~\ref{fig:saslaw-record-info} illustrates the recording configuration during a conversation between two participants.
SaSLaW records what a participant speaks, listens to, and watches synchronously across two participants. Speech recordings can be utilized for constructing TTS models to generate natural speech incorporating human-like auditory and visual information. 

\vspace{-1mm}
\subsection{Recording Configuration and Procedure} \vspace{-1mm}
\textbf{Participant equipment.} Two participants engage in conversation in a single indoor space, referred to as the recording room. They sat facing each other across a table.
The distance between the two participants is set within $1.5$ to $3\, \mathrm{m}$ and is not strictly controlled.
Participants are equipped with a close-talking microphone\footnote{\fontsize{6pt}{6pt}\selectfont \url{https://www.shure.com/en-US/products/microphones/pga31}}, ear-mounted binaural microphone\footnote{\fontsize{6pt}{6pt}\selectfont \url{https://soundprofessionals.com/product/MS-EHB-2/}}, and head-mounted camera\footnote{\fontsize{6pt}{6pt}\selectfont\url{https://ordro.online/en-jp/products/camcamcorder-ep8}} on their heads, as illustrated in Figure \ref{fig:saslaw-record-info}. 
Each device corresponds to recording what they speak, hear, and see. The two microphones record at a sampling frequency of $44.1 \, \mathrm{kHz}$, while the head-mounted camera records at a frame rate of $30$ fps. All six sensors record information synchronously.
During the recording, variations in speech volume due to the audio environment are expected. To capture these variations, the gain for each participant's microphones is fixed throughout all recordings.

\textbf{Environmental noise.} To simulate various real environments with noise, eight loudspeakers are positioned as shown in Figure \ref{fig:saslaw-record-info}, covering the area around the participants. Each loudspeaker plays a different segment of the same environmental noise, simulating diffusive environmental noise in the real environment.
The real-environmental-noise data are derived from a subset of the DEMAND dataset~\cite{demand}. The type and power of the noise played from loudspeakers are altered after a certain number of conversation recordings. Before recording the conversation, the ambient noise level in dB is measured at the center of the two participants using a noise meter\footnote{\fontsize{6pt}{6pt}\selectfont\url{https://www.sanwa.co.jp/product/syohin?code=CHE-SD1}}.

\textbf{Conversation content.} The two participants are instructed on the theme and roles (e.g. \textit{sightseeing, the guide and tourist}) and tasks.
During the recording, the two participants engage in improvisational conversation consisting of five to eight turns, following provided instructions.

\textbf{Annotation as TTS corpus.} To use the SaSLaW for TTS use, we automatically segment close-talking microphone voices using pyannote.audio~\cite{Bredin23} into utterances and transcribe them into texts using whisper~\cite{whisper}, followed by manual correction.

\vspace{-1mm}
\subsection{Data Collection for Reproducible Evaluation} \vspace{-1mm}
\label{subsec:reproducible-evaluation}
In subjective evaluations, evaluators should assess the plausibility within audio environments based on what they would listen to at the listener position while models output speech via sonic transmission, rather than synthetic speech itself.
Therefore, we collect supplementary data following the previous evaluation methodology~\cite{rennies20interspeech}. SaSLaW records impulse responses from talker to listener, positioned as shown in Figure~\ref{fig:saslaw-record-info}, and ambient noise-only audio in the listener's position using the ear-mounted binaural microphone.
Impulse responses are recorded for various distances between participants. 
Synthetic utterance samples are convolved with a certain impulse response and added with recorded noise-only audio. Then we acquire the evaluation samples simulating what listeners would hear.

\vspace{-1mm}
\section{Corpus Analysis} \vspace{-1mm}
\label{sec:corpus-analysis}
Four pairs of Japanese participants\footnote{There are three male-male pairs and one female-female pair.} engaged in the recordings of spontaneous dialogues, as outlined in Section \ref{sec:corpus-construction}. This section reports on the analysis of two pairs' SaSLaW recordings. One pair consisted of two male participants (named as \textit{spk01}, \textit{spk02}), while the other pair consisted of two female (named as \textit{spk05}, \textit{spk06}).
The total utterance length of recorded speech was approximately 30 minutes on average.

\vspace{-1mm}
\subsection{Purpose and Procedures of Corpus Analysis} \vspace{-1mm}
\label{subsec:corpus-analysis-procedures}
This analysis examined how spontaneous speech changed depending on the sound pressure level of environmental noise and inter-speaker coordination. Note that we did not investigate the distance between participants.

First, each conversation was assigned a label about environmental noise levels (\textit{env-label}) such as ``noisy'' (high environmental noise level), ``moderate'' (moderate), or ``quiet'' (low). The label annotation was based on the sound pressure level of environmental noise measured in the conversation recording. 
Next, the root mean square of amplitudes (RMS), $F_0$, $F_1$ (first formant) frequency, and spectral tilt~\cite{sato2023asru} were calculated for voiced frames of utterances as prosodic features. $F_0$ was computed and voiced frames were detected using harvest~\cite{morise17b_interspeech}.
The $F_1$ frequency was computed through Praat~\cite{praat}.
Spectral tilt was computed following previous research~\cite{sato2023asru}, with a filter bank from $0.25$ to $8 \, \mathrm{kHz}$.
All the statistical tests below were conducted at a significance level of $p = 0.05$.

\vspace{-1mm}
\subsection{Analysis Results and Discussion} 
\begin{figure}
    \centering
    \includegraphics[width=1.05\linewidth]{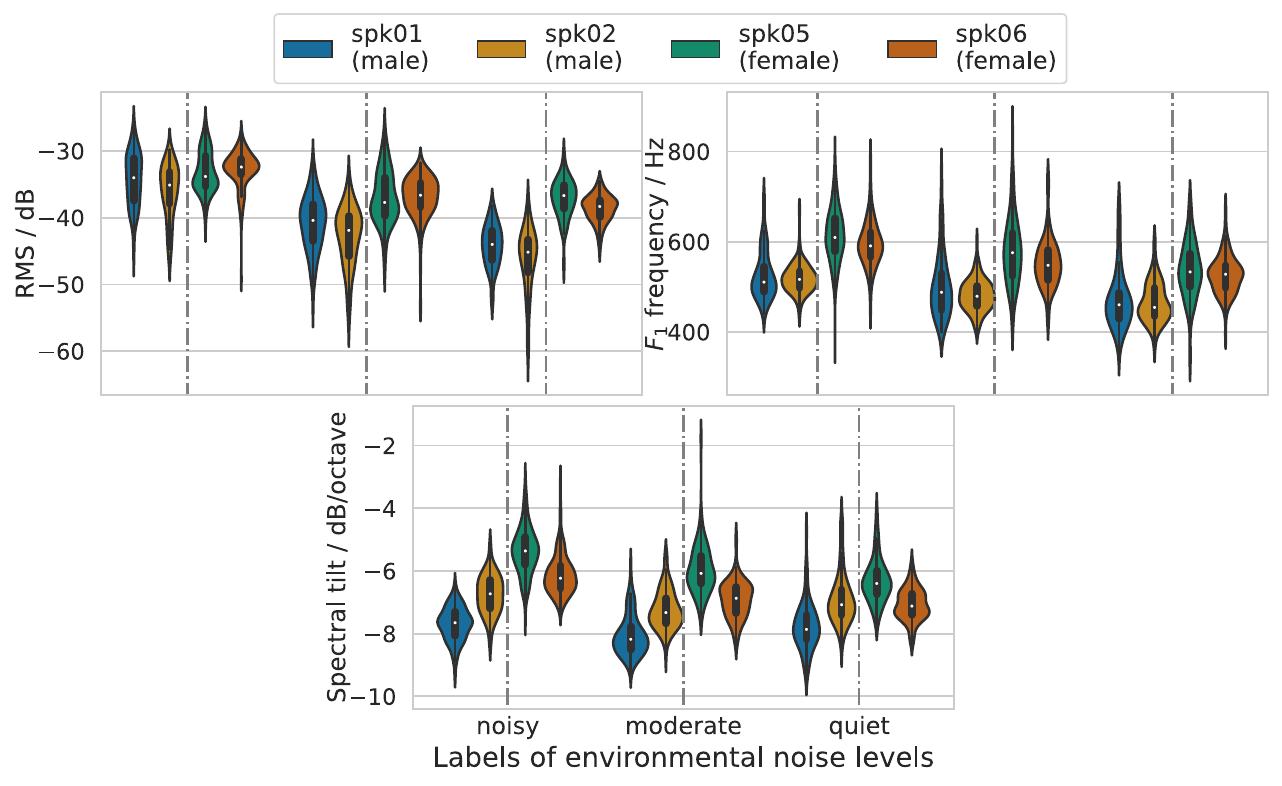}
    \vspace{-6mm}
    \caption{Feature distributions of each speaker.}
    \label{fig:corpus-info}
\end{figure}

\vspace{-1mm}
\label{subsec:analysis-result}
\textbf{Noise.} Figure~\ref{fig:corpus-info} illustrates the distributions of RMS, $F_1$ frequencies and spectral tilts for each speaker.
It is shown that the average $F_1$ frequency significantly increased from ``quiet'' to ``noisy'' env-labels for all the speakers. 
The average of RMS showed a significant rise from ``quiet'' to ``noisy,'' for all the speakers except for spk05.
The average spectral tilt presented a significant increase for female speakers from ``quiet'' to ``noisy,'' whereas a different trend was observed for male.

These increases
corresponded to the result reported in the previous study~\cite{alghamdi2018jasa}.
Also, the result suggests that while some features share characteristics among speakers in adapting to varying levels of environmental noise, others do not exhibit such commonality.
It indicates that solely relying on rule-based methods with signal processing makes it difficult to accurately simulate Lombard speech adapting to audio environments.

\textbf{Interlocutors.} Table~\ref{tab:corr-inter-speakers} shows the inter-speaker correlation coefficients of utterance RMS and $F_0$. 
\begin{table}[tb]
    \centering
    \caption{The correlation coefficients of the RMS and $F_0$ between the speaker's utterance and the interlocutor's last utterance before the speaker's. Significant correlations are displayed \textbf{bold}.}
    \footnotesize
    \vspace{-3mm}
    \begin{tabular}{rccc||ccc}
         &  \multicolumn{3}{c}{\textit{spk01-spk02}} & \multicolumn{3}{c}{\textit{spk06-spk05}} \\\hline\hline
         & noisy & moderate & quiet & noisy & moderate & quiet\\
         RMS & $\mathbf{0.68}$ & $\mathbf{0.65}$ & 0.07 & $\mathbf{0.24}$ & $\mathbf{0.23}$ & 0.00 \\
         $F_0$ & $\mathbf{0.47}$ & 0.10 & 0.21 & $\mathbf{0.23}$ & $\mathbf{0.35}$ & 0.15\\\hline
    \end{tabular}
    \label{tab:corr-inter-speakers}
\end{table}
The result indicates that in moderate and noisy environments, the target utterance features are significantly correlated with those of the last interlocutor's utterances. 
Adaptations to harsh listening environments may evoke this correlation enhancement from quiet to moderate, noisy. This discussion is akin to the report about the relationship between conversation excitement and entrainment in a previous study~\cite{nishimura09jpsj}.
These results and discussion also suggest that SaSLaW is suitable for modeling the comprehensive consideration of and adaptation to audio environments, including noise and interlocutor factors.

\vspace{-1mm}
\section{Environment-adaptive TTS Experiment}\vspace{-1mm}
We utilized spk01 and spk06 data to construct single-speaker EA-TTS models for each speaker. We evaluated the plausibility of synthetic speech in various audio environments.

\vspace{-1mm}
\subsection{Experimental Conditions}
\begin{figure}[t]
    \hspace{-5mm}
    \centering
    \includegraphics[width=1.05\linewidth]{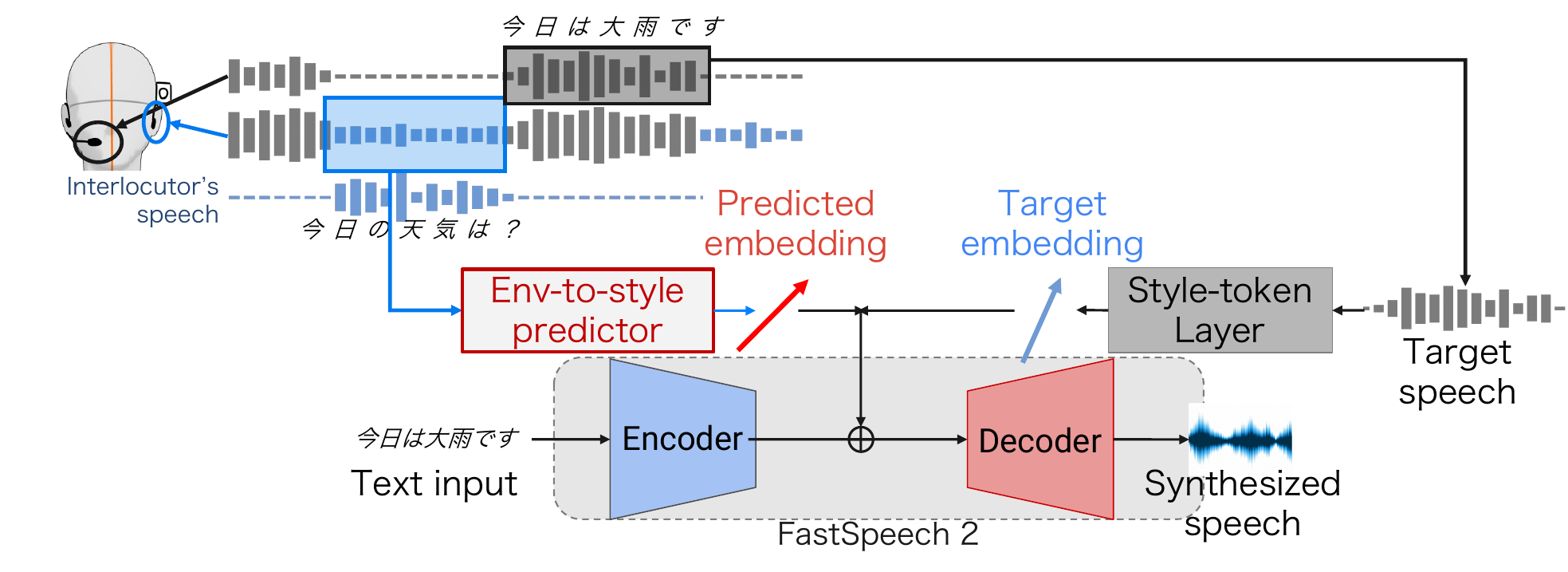}
    \vspace{-2mm}
    \caption{Diagram of EA-TTS model. This model predicts the style vector extracted from the target utterance in training.}
    \label{fig:proposed-ea-tts}
\end{figure}

\vspace{-1mm}
We compared three neural EA-TTS models. All of the models basically employed open-sourced FastSpeech 2~\cite{ren2022fastspeech} and HiFi--GAN~\cite{hifigan}\footnote{\scriptsize\url{https://github.com/Wataru-Nakata/FastSpeech2-JSUT}. We also followed its hyperparameter settings.}. 
The details of these models are as follows.
\begin{itemize}
    \item \textbf{FS2}: basic FastSpeech 2 fine-tuned on SaSLaW speech data.
    \item \textbf{FS2-predsty}: EA-TTS model fine-tuned on SaSLaW speech itself and auditory input during the interlocutor's last turn preceding that speech (mixed with background noises). 
    \item \textbf{FS2-predsty-ptrn}: FS2-predsty, further pre-trained on pseudo-environment-adaptive data and fine-tuned on SaSLaW speech and hearing-audio data.
\end{itemize}
FS2 and \{FS2-predsty, FS2-predsty-ptrn\} models have 35M and 61M parameters, respectively. FS2 and FS2-predsty were pre-trained on the JSUT~\cite{jsut}, an existing TTS corpus.
Figure~\ref{fig:proposed-ea-tts} illustrates the EA-TTS model applied to FS2-predsty and FS2-predsty-ptrn. EA-TTS incorporates a style-token layer and Env-to-style predictor into the original FastSpeech 2. The style-token layer uses global style token~\cite{wang2018icml}, which extracts a style vector with a fixed length from an utterance. The Env-to-style predictor consists of four trainable convolution layers and an energy extractor, which predict the style vector from hearing audio. The objective $L$ is defined as $L = L_{\mathrm{TTS}} + L_{\mathrm{sty}}$, where $L_{\mathrm{TTS}}$ denotes the objective of FastSpeech 2 and $L_{\mathrm{sty}}$ denotes the L1 loss between style vectors.

\textbf{Pseudo-data training}. We created a pseudo-environment-adaptive dataset by using a TTS corpus and noise dataset through signal processing. While utterances manipulated in a signal-processing manner deviate in features from real Lombard speech, these utterances can effectively pretrain EA-TTS models due to their scalability. We assign each utterance an arbitrary noise signal at a varying level then increase its spectral tilt to enhance intelligibility within the corresponding noise. We use JSUT as the TTS corpus and DEMAND as the noise dataset.

\textbf{Training conditions}. We split SaSLaW audio recordings into train and test sets for each speaker's data. There was no overlap in the environmental noise contained in the hearing audio between the sets. The test set covered all env-labels assigned to speech. Finally, we split the recordings of spk01 into 299/49 and spk06 into 443/64 train/test utterances. All three EA-TTS models were trained on a single NVIDIA GeForce RTX 4090 GPU.
They were pre-trained for 900k ($\leq$ three days) and fine-tuned on single-speaker recordings of SaSLaW for 100k steps ($\leq$ 12 hours).

\subsection{Objective Evaluation}
\begin{figure}[t]
    \centering
    \includegraphics[width=\linewidth]{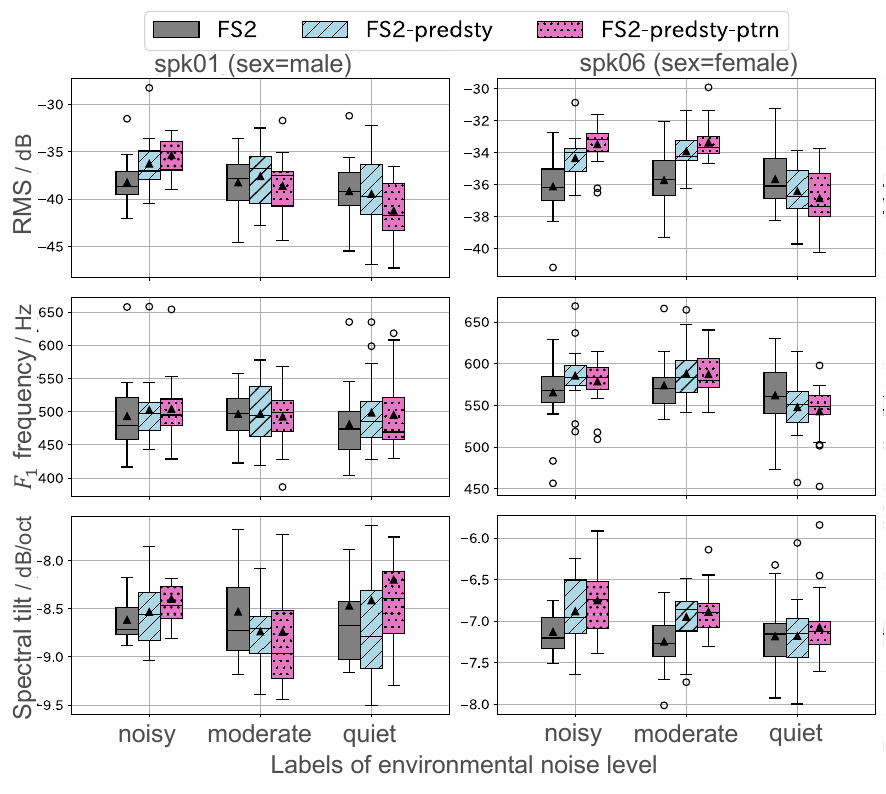}
    \vspace{-5mm}
    \caption{Feature distributions of synthetic speech. Solid triangles indicate the average of distributions, and black lines indicate centroid.}
    \label{fig:objective-result}
\end{figure}
For objective evaluation, we computed the prosodic features outlined in Section \ref{subsec:corpus-analysis-procedures} for the synthetic speech.
We investigated whether the distribution changed across different env-labels, consistent with the analysis presented in Figure \ref{fig:corpus-info}. 

Figure \ref{fig:objective-result} illustrates the prosodic feature distributions of synthetic speech. FS2 showed no significant differences across env-labels.
For FS2-predsty and FS2-predsty-ptrn, the synthetic speech of spk01 showed a significant increase only for RMS from ``quiet'' to ``noisy''. For spk06, the synthesized speech exhibited a significant increase from ``quiet'' to ``noisy'' for all features except between ``moderate'' and ``noisy'' for spectral tilt. 
These results indicate that the Env-to-style predictor enabled the generation of speech with characteristics adapted to audio environments.

\vspace{-1mm}
\subsection{Subjective Evaluation} \vspace{-1mm}
\begin{figure}[t]
    \centering
    \includegraphics[width=\linewidth]{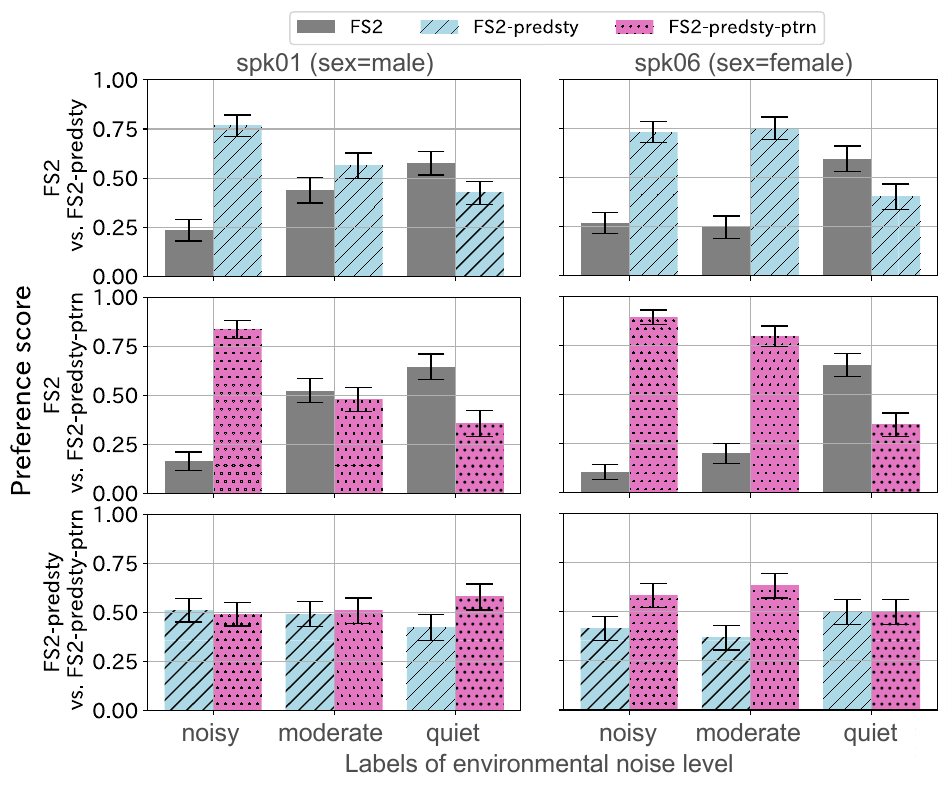}
    \vspace{-5mm}
    \caption{Preference score of each model pair. Error bars denote 95\% confidence intervals.}
    \label{fig:ab-result}
\end{figure}
We conducted an AB preference test to compare the plausibility of the evaluation speech samples within surrounding noises. Each synthetic utterance was processed as described in Section \ref{subsec:reproducible-evaluation} to serve as the evaluation utterance. Evaluators selected which utterance sounded more plausible in environmental noises through the listening experiment. The evaluators were provided with a combined criterion of naturalness and intelligibility as supplementary instruction. For each model-pair configuration, 72 evaluators were recruited and 720 responses were collected via Lancers\footnote{\url{https://www.lancers.jp}}. 
Figure \ref{fig:ab-result} illustrates the env-label-wise preference scores for each model-pair and speaker. 

The results indicate that for the ``noisy'' label, both FS2-predsty and FS2-predsty-ptrn significantly outperformed FS2 in preference scores for both speakers, while FS2 significantly surpassed the other two for the ``quiet'' label. This result suggests that the EA-TTS models with the Env-to-style predictor successfully adapted to noisy surroundings. However, this adaptation to ``quiet'' environments degraded the plausibility of synthetic speech compared with FS2-synthesized speech, which had averaged prosodic features and gained intelligibility.

Figure \ref{fig:ab-result} shows that the preference scores of FS2-predsty-ptrn were equal to or significantly higher than those of FS2-predsty. 
This suggests that pseudo-data pre-training improved the performance.

\vspace{-1mm}
\section{Conclusion} \vspace{-1mm}
We introduced SaSLaW, a novel speech corpus for generative tasks with synchronized audio and visual first-person recordings. We described the methodology of constructing SaSLaW and analyzed the recordings to confirm human speech's adaptations to audio environments. The experimental results indicate that SaSLaW enables the construction of environment-adaptive TTS models by using the auditory perception of the target speaker as input, successfully producing plausible speech tailored to diverse audio environments. This work does not explore the analysis and modeling of speech adaptation to visual information, which can be investigated for further work.

\section{Acknowledgements}

Part of this work was supported by JSPS KAKENHI Grant Number 23H03418 and 22H03639, Moonshot R\&D Grant Number JPMJPS2011, and JST FOREST JPMJFR226V.
The authors also thank Aya Watanabe for her support in designing the figures for this paper.


\bibliographystyle{IEEEtran}
\bibliography{mybib}

\end{document}